\documentclass{aa}  

\usepackage{xcolor}
\usepackage{graphicx}
\usepackage[varg]{txfonts}
\begin{document} 

\title{Detection of volatiles undergoing sublimation from 67P/Churyumov-Gerasimenko coma particles using ROSINA/COPS}
\subtitle{I. The ram gauge}

\author{B. Pestoni\inst{1} \and K. Altwegg\inst{1} \and H. Balsiger\inst{1} \and N. H{\"a}nni\inst{1} \and M. Rubin\inst{1} \and I. Schroeder\inst{1} \and M. Schuhmann\inst{1} \and S. Wampfler\inst{2}}

\institute{Physics Institute, Space Research \& Planetary Sciences, University of Bern, Sidlerstrasse 5, 3012 Bern, Switzerland\\ \email{boris.pestoni@space.unibe.ch}
\and
Center for Space and Habitability, University of Bern, Gesellschaftsstrasse 6, 3012 Bern, Switzerland}

\date{Received 9 August 2020 / Accepted 30 November 2020}


\abstract
{The ESA Rosetta mission has allowed for an extensive in situ study of the comet 67P/Churyumov-Gerasimenko. In measurements performed by the ram gauge of the COmet Pressure Sensor (COPS), observed features are seen to deviate from the nominal ram gauge signal. This  effect is attributable to the sublimation of the volatile fraction of cometary icy particles containing volatiles and refractories.}
{The objective of this work is to investigate the volatile content of icy particles that enter the COPS ram gauge.}
{We inspected the ram gauge measurements to search for features associated with the sublimation of the volatile component of cometary particles impacting the instrument. All the sublimation features with a high-enough signal-to-noise ratio were modelled by fitting one or more exponential decay functions. The parameters of these fits were used to categorise different compositions of the sublimating component.}
{Based on features that are attributable to ice sublimation, we infer the detection of 73 icy particles containing volatiles. Of these, 25 detections have enough volatile content for an in-depth study. From the values of the exponential decay constants, we classified the 25 inferred icy particles into three types, interpreted as different volatile compositions, which are possibly further complicated by their differing morphologies. The available data do not give any indication as to which molecules compose the different types. Nevertheless, we can estimate the total volume of volatiles, which is expressed as the diameter of an equivalent sphere of water (density of 1 g cm$^{-3}$). This result was found to be on the order of hundreds of nanometres.}
{}


\titlerunning{Detection of volatiles from 67P coma particles using COPS-RG}
\maketitle


\section{Introduction}\label{sec:introduction}
The Rosetta mission was the first opportunity to carry out extensive in situ analyses of the nucleus of a comet and its respective surroundings \citep[][]{Taylor_et_al_2017}. The target of the mission, namely, the Jupiter-family comet 67P/Churyumov-Gerasimenko (hereafter, 67P), was investigated with a suite of 11 instruments on the Rosetta orbiter \citep[][]{Vallat_et_al_2017} and 10 additional instruments on Philae, the lander module \citep[][]{Boehnhardt_et_al_2017}. Rosetta contained multiple instruments devoted to the study of dust particles coming from the nucleus of 67P. One such instrument, namely, the Grain Impact Analyzer and Dust Accumulator \citep[GIADA,][]{Colangeli_2007}, measured the momentum distribution and mass of dust grains from 67P. The particles collected by GIADA can be organised into two groups: 1) compact particles with sizes in the range from 0.03 mm to 1 mm; and 2) fluffy aggregates (made of sub-micron grains) with sizes from 0.2 mm to 2.5 mm \citep{Rotundi_et_al_2015,Fulle_2015}. Another dedicated dust instrument was the Micro-Imaging Dust Analysis System \citep[MIDAS,][]{Riedler_2007}, which is an atomic force microscope that resolved dust particles in the size domain of a micron. \citet{Mannel_et_al_2019} found that refractory particles gathered by MIDAS are composed of grains having a size of about one hundred nanometres. Yet another instrument was the COmetary Secondary Ion Mass Analyser \citep[COSIMA,][]{kissel_2007}, which collected, took pictures and analysed dust particles of cometary origin by secondary ion mass spectrometry. \citet{bardyn_2017} obtained the elemental composition of dust by analysing some of the tens of thousands of particles collected by this instrument.

This work focuses on the COmet Pressure Sensor (COPS), which is one of the three sensors of the Rosetta Orbiter Spectrometer for Ion and Neutral Analysis \citep[ROSINA,][]{Balsiger_et_al_2007}.

\begin{figure}
\centering
\includegraphics[width=\hsize]{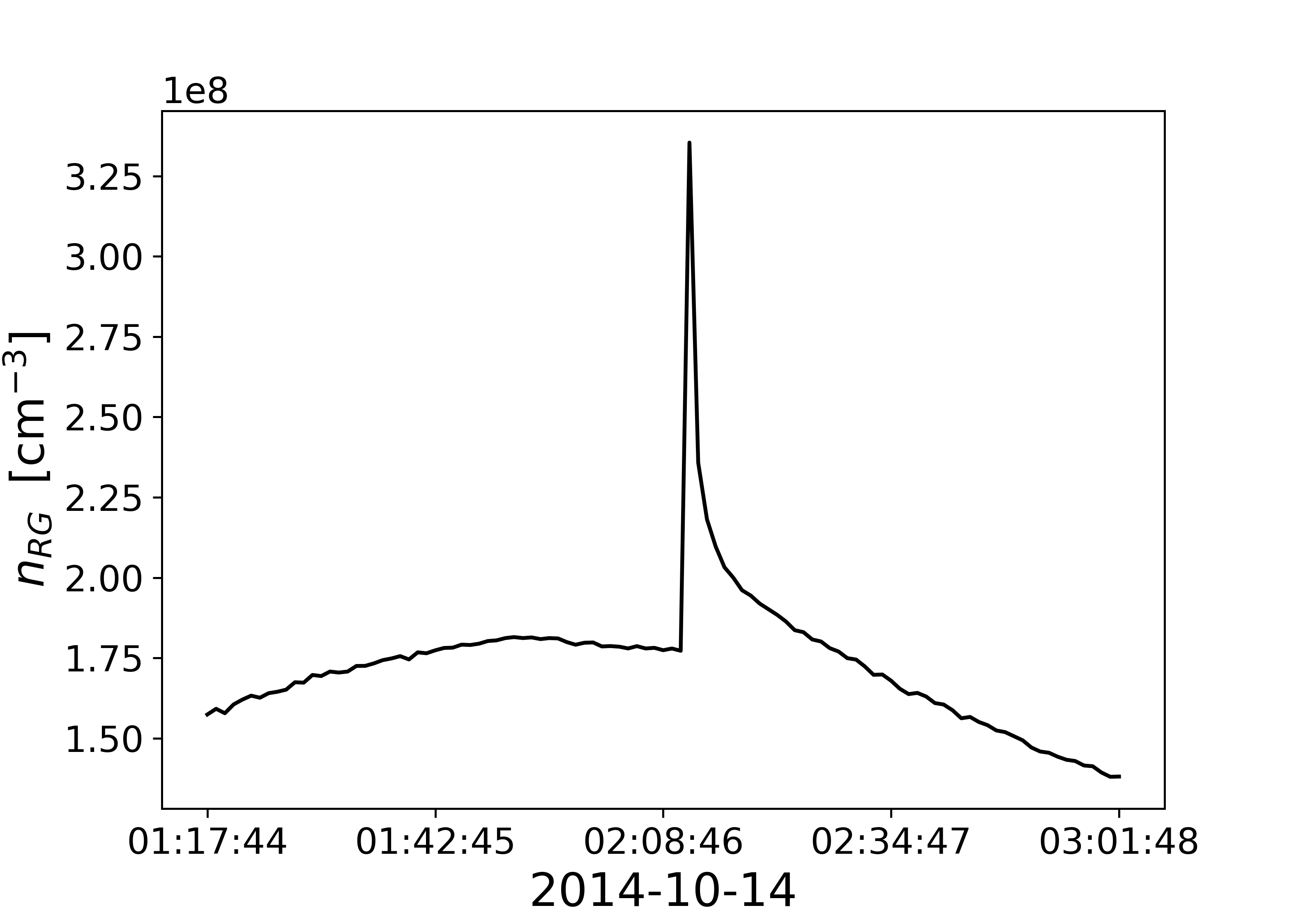}
\caption{Density measured by the COPS ram gauge (RG) on 14 October 2014 between 01:17:44 and 03:01:48. The graph shows a regular nominal RG signal that is interspersed with a sudden increase at 02:12:46, followed by a gradual decrease back to nominal RG levels lasting about 20 min.}
\label{fig:motivation}
\end{figure}

The main scientific goal behind COPS is to determine the cometary gas density in the coma of 67P. Although it has not been designed specifically to detect cometary particles, COPS is also able to measure the sublimation products of their volatiles. This fact can be seen clearly in Fig.~\ref{fig:motivation}, where the density measured by the COPS ram gauge (henceforth, RG) in the early hours of 14 October 2014 is shown. 
A sudden increase in density is observed at a time when the underlying cometary gas density has an almost smooth evolution. An abrupt increase in the measured density implies there must necessarily have been a brief additional source of gas located inside the RG. Sudden decreases, on the other hand, have not been observed with COPS. As we go on to explain in Sect.~\ref{subsec:features_and_dust}, such a feature arises when the volatile part of an icy particle sublimates after entering the RG. We define an icy particle as a dust particle consisting of a condensed volatile component (i.e.\ material with a low condensation point) and, potentially, also a refractory component (i.e.\ material with a high condensation temperature) that is invisible to COPS and, hence, our analysis as well. Icy particles have been found to contribute significantly to the water production rate for the comet Hartley 2 \citep[e.g.\ ][]{Protopapa_et_al_2014}. For 67P, an upper limit on the contribution of sublimating icy particles on this production rate was derived \citep[50\%,][]{Biver_et_al_2019}. 
Ice is present in cometary particles either in the form of grains with sizes of about one hundred of nanometres or as shells embedding refractory grains of close to one hundred nanometres \citep{Guttler_et_al_2019}. Despite the fact that was equipped with the right resolution for measuring single isolated submicron grains, MIDAS was not successful in detecting any of these particles.

Due to to the aforementioned connection between an increase in measured RG density and icy particles, the presence of the latter can be identified by the RG after subtraction of the nominal RG signal (i.e. the nominal density background caused by gaseous coma species that is measured by the RG). This must be done following the careful removal of features associated with spacecraft background effects, such as thruster firings, offset measurements, and (large) slews \citep[][]{Schlappi_et_al_2010}, particularly as the latter can be wrongly interpreted as a consequence of the sublimation of the volatiles of an icy particle. 

This paper is organised as follows: Sect.~\ref{sec:methods} introduces the methodological framework, Sect.~\ref{sec:results} presents the results, and Sect.~\ref{sec:conclusion} discusses the limitations of the findings. We conclude with a summary and an outlook on possible future works.


\section{Methods}\label{sec:methods}

This section is organised into four main parts that describe: the relationship between increases in density observed by the RG and the presence of icy particles inside the instrument (Sect.~\ref{subsec:features_and_dust}), a description of the aspects of the ram gauge relevant to this work (Sect.~\ref{subsec:cops_ram_gauge_and_sizes}), the identification of features caused by the sublimation of the volatile component of cometary icy particles (Sect.~\ref{subsec:peaks_identification}), and the procedure required to analyse the observations (Sect.~\ref{subsec:analysis_procedure}).

\subsection{Features produced by icy particles}
\label{subsec:features_and_dust}
In this section, we explaine why we assign the RG COPS features to the sublimation of the volatile component of icy particles.

On 5 September 2016, several instruments on board the Rosetta spacecraft –- GIADA, the Optical, Spectroscopic, and Infrared Remote Imaging System \citep[OSIRIS,][]{thomas_1998}, and ROSINA -- reported a sharp increase in dust activity \citep[][]{Altwegg_et_al_2017,della_corte_2019}. In particular, the dust impact on 5 September 2016 was clearly registered by all GIADA sensors \citep{della_corte_2019} of one magnitude higher than during the 19 February 2016 outburst \citep{Grun_et_al_2016}. However, even instruments dedicated to the measurement of coma gases were directly affected by this outburst. \citet{Altwegg_et_al_2017} reported a temporary physical blockage of the ion source of the ROSINA Double Focusing Mass Spectrometer \citep[DFMS, another ROSINA instrument,][]{Balsiger_et_al_2007}. Additionally, the star trackers experienced difficulties in carrying out their respective functions. These observations are consistent with the large quantity of dust from 67P impacting the Rosetta spacecraft on 5 September 2016.

Over the course of this event, the distance between the comet and the Rosetta spacecraft (cometocentric distance) was about 5 km and the velocity of dust particles, as measured by GIADA, was $1-10$ ms$^{-1}$ \citep{Della_Corte_2015}. Consequently, the dust particles would have required less than ten minutes to traverse the distance between 67P's surface and the spacecraft. This was too short a time interval for most of the volatiles and semivolatiles to reach a sufficiently high temperature to become sublimated before reaching Rosetta \citep{Altwegg_et_al_2017,Altwegg_et_al_2020}. This means that if there were volatiles in the particles ejected during this event, they could have been measured by instruments on board the spacecraft. Some of these molecules were indeed observed by DFMS, providing evidence that volatiles and semivolatiles were present in the solid material expelled during this episode. The study of the DFMS measurements led to the identification of a wide range of species \citep[][]{Altwegg_et_al_2017,Schuhmann_et_al_2019,Altwegg_et_al_2020}. The identified species were found to be compatible both with the measurements made on 12 November 2014 by the two mass spectrometers: Ptolemy \citep[][]{Morse_et_al_2009} and the Cometary Sampling and Composition \citep[COSAC,][]{Goesmann_et_al_2007} on the surface of 67P \citep[][]{Goesmann_et_al_2015,Wright_et_al_2015} and with the measurements between August and September 2014 carried out by the Visible and Infrared Thermal Imaging Spectrometer \citep[VIRTIS,][]{Coradini_et_al_2007,Capaccioni_et_al_2015}, an imaging spectrometer on board Rosetta.

The DFMS was not the only ROSINA instrument that performed measurements during the outburst on 5 September 2016, as COPS was also operational. The Reflectron Time-of-Flight mass spectrometer \citep[RTOF,][]{Balsiger_et_al_2007}, however, was switched off due to power restrictions. The COPS RG measured several sudden increases in density such as the one in Fig.~\ref{fig:motivation}. The concurrence of these peculiar features with observations by GIADA, OSIRIS, and DFMS confirms that they are generated by the dust coming from 67P. Since COPS is a pressure sensor (see Sect.~\ref{subsec:cops_ram_gauge_and_sizes} for a description of its function), it can only detect dust when the volatile part of an icy particle is sublimated.
This is applicable not only to the outburst of 5 September 2016, but also at any other time during the mission.
Therefore, whenever a feature similar to the one displayed in Fig.~\ref{fig:motivation} is observed by COPS (and spacecraft background changes due to e.g. slews and thruster firings can be excluded) indicates that an icy particle has entered COPS and that its volatile content has been sublimated.

\subsection{Ram gauge measuring principle}
\label{subsec:cops_ram_gauge_and_sizes}
\begin{figure}
\centering
\includegraphics[width=0.25\textwidth]{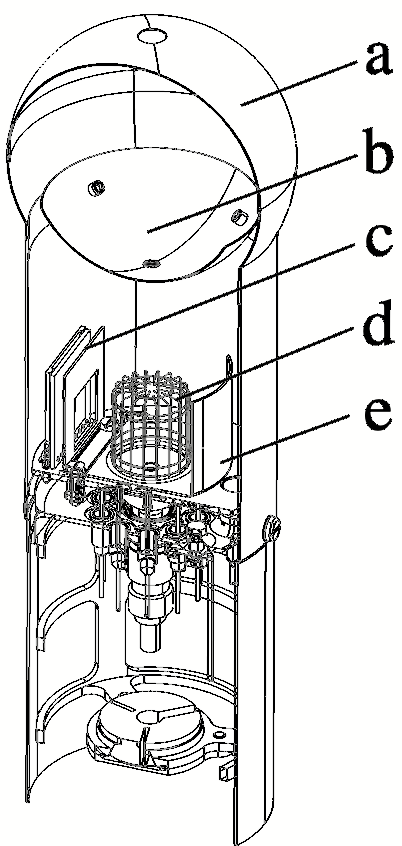}
\caption{Sketch of the COPS ram gauge. The main components are: (a) a 60 mm diameter spherical equilibrium cavity with a round 6 mm opening pointing towards 67P; (b) an inner shield; (c) a microtip field emitter device; (d) a 19 mm in height and 8 mm in radius anode at +180 V; and (e) an electron repeller.}
\label{fig:ram_gauge}
\end{figure}

The COPS is composed of two density gauges: the nude gauge, which measures the total neutral density, and the RG, which measures the ram pressure (proportional to the cometary gas flux). This study is focused on the RG, illustrated in Fig.~\ref{fig:ram_gauge}.

Gas originating from the comet enters the instrument through the 6 mm diameter circular opening and remains confined within the spherical cavity (a). Since the opening is much smaller than the cavity, the possibility of escape for gas molecules is strongly reduced. Due to numerous collisions, the gas is first isotropised and thermalised at the temperature of the spherical cavity; this temperature is actively measured by the instrument. The ionisation region, which is separated by an inner shield (b) and also by a grid that repels ions from outside (d), operates according to the extractor-type ionisation gauge principle \citep[][]{Redhead_1966}. The working principle is the same as for the nude gauge (see \citet{Balsiger_et_al_2007} and \citet{Graf_et_al_2008} for a more detailed explanation): the ionised particles are accelerated towards a base plate, focused by a reflector, and finally collected by an electrometer. The biggest difference between the two gauges is the manner employed to ionise the gas. Unlike the nude gauge, the RG does not use a hot filament. In order to keep the walls of the spherical cavity and the boom containing the ionisation region at a moderate temperature (the surrounding coma is much colder than the typical 0$^{\circ}$ C of the RG), it uses a microtip field emitter (c), which is a type of cold electron source. Opposite the microtip, an electron repeller (e) sends electrons back to the ionisation region, hence increasing the ionisation efficiency of the gauge. 

Combining the aforementioned measuring principle with the correlation between increases in density and the presence of icy particles inside the RG (presented in Sect.~\ref{subsec:features_and_dust}), it is possible to determine when there is a dust particle with a volatile component inside the instrument.

\subsection{Feature identification}
\label{subsec:peaks_identification}
In Sect.~\ref{sec:introduction}, we remark on the fact that there are several factors that complicate the identification of features from the measurements of the RG. The main difficulties encountered in the analysis are: (1) the high nominal RG signal; (2) the noise in the measurements; (3) the short duration of the feature; and (4) the offset measurements. These are discussed in detail in the following:

The density measured by the RG is generally high (the average of the measurements is $\sim10^8\,\mathrm{cm^{-3}}$). This means that it is arduous to resolve small features as they blend into the nominal RG signal. 

Additionally, the RG exhibits a very noisy behaviour, making it difficult to discern features. Thus, the best conditions for identifying features are when the density is relatively uniform or the amount of volatiles in the icy particle leads to a feature dominating the nominal RG signal.

Another issue that complicates the identification of icy particles detected by the RG is the occasional scarcity of data points that make up the features generated by the sublimation of the volatile content of these particles. This can happen either if the feature is barely above the nominal RG signal as previously mentioned or if the volatile part of the icy particle undergoes sublimation particularly quickly, resulting in very sparse instances of data acquisition of the event. 

\begin{figure}
\centering
\includegraphics[width=\hsize]{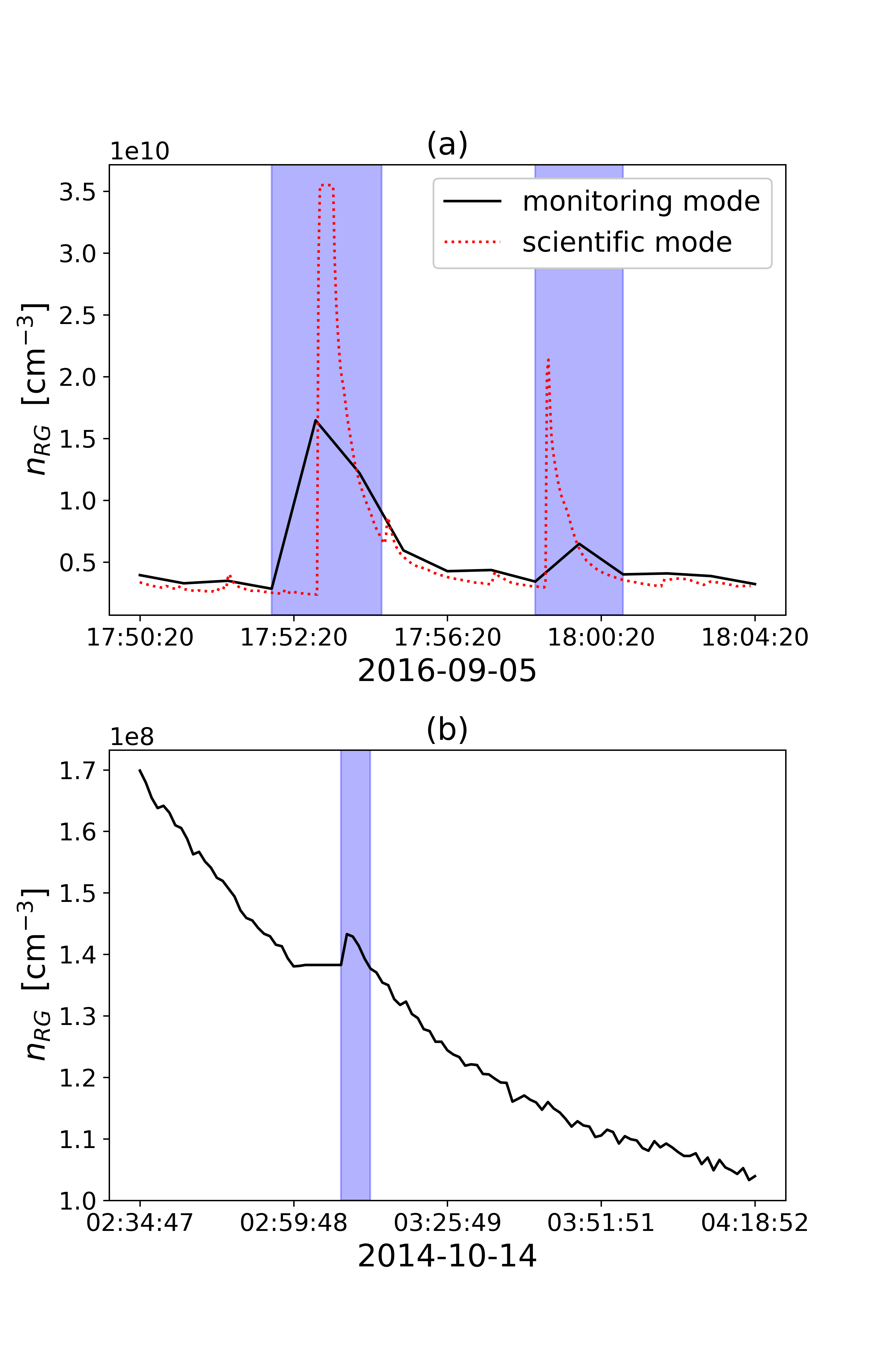}
\caption{Examples of two of the difficulties that can be encountered during the extraction of features generated by the sublimation of volatiles in icy particles. \textit{Panel a}: density measured by the RG (black solid line for the monitoring mode, red dotted line for the science mode) on 26 October 2014. Two density peaks are observable which, although attributable to the sublimation of volatiles in icy particles, are not resolved enough to display the tail in the monitoring mode. In this particular case, the science mode was active, so the tail of both features is visible when considering the more refined mode. It should be noted that the upper part of the first feature measured using the science mode is flat, meaning that the RG has reached saturation. This aspect is discussed in Sect.~\ref{subsec:analysis_procedure}.
\textit{Panel b}: density measured by the RG on 14 October 2014. This plot shows an example of a feature that could erroneously be interpreted as the presence of an icy particle inside the RG. However, this feature can be confidently attributed to an offset measurement because just before the increase in density, there is a plateau lasting eight minutes, during which the measured values are exactly the same.}
\label{fig:difficulties}
\end{figure}

The COPS has two operating modes: the most frequently used one is called ``monitoring mode'' and stores a value every minute, whereas the rarely used ``scientific mode'' tracks a value every two seconds \citep[][]{Balsiger_et_al_2007}.
As we show in Sect.~\ref{sec:results}, all the features obtained from the scientific mode are fully time-resolved: there is an abrupt increase in density, immediately followed by an exponential decay (hereafter, identified as the tail) until it reaches values similar to the ones before the increase. Sometimes, during the measurement of a feature using the scientific mode, the instrument reaches saturation (see the largest red peak in Fig.~\ref{fig:difficulties}a). This introduces an error in the analysis caused by the difficulty in establishing with certainty the starting time of the tail.

Measurements taken in the monitoring mode were analysed after the ones in science mode had been exhausted. Some of these had clear features (as the one shown in Fig.~\ref{fig:motivation}), while others were less evident due to the scarcity of data points. The reason for this is that when we have a time resolution of only one minute, it is possible that the volatiles of some icy particles are sublimated in the interval between two consecutive instances of data acquisition, so the tail is not visible. In Fig.~\ref{fig:difficulties}a, an example of single point peaks measured using the monitoring mode (black solid line) is shown. In this specific case, the science mode was also operational (red dotted line) and, therefore, it is possible to retrieve the sublimation data from the latter. However, science mode measurements were rarely available and, therefore, in the majority of cases where the features in monitoring mode were not well-resolved, it was not possible to extract information from the tail, such as the amount of volatiles in the icy particle.

Finally, it should be noted that not all features that deviate from the usual behaviour of the nominal RG signal are associated with icy particles. There are similar features following offset measurements, as can be seen from the example in Fig.~\ref{fig:difficulties}b. An offset measurement is a series of measurements made roughly every 24 hours by switching off the electron emission from the microtip field emitter. The purpose of this procedure is to reset the offset that the electrometer could have accumulated during nominal operation. Nonetheless, this kind of feature is easily identifiable due to the presence of a plateau, composed of at least five data points with exactly the same value, just before the density increase. This unique characteristic allows for the exclusion of these data from the analysis.

\subsection{Analysis procedure}
\label{subsec:analysis_procedure}
We analysed features showing a sufficiently time-resolved tail (i.e.\ at least five data points) for the different volatile components and for the total volume of the volatile part of the icy particles. The analysis procedure, described in detail below, is based on several assumptions and the density measured by COPS is normalised to $\mathrm{N_2}$, hence, there are correction factors to account for a $\mathrm{H_2O}, \mathrm{CO}$, and $\mathrm{CO_2}$ dominated coma \citep[][]{Gasc_et_al_2017}.

\begin{figure}
\centering
\includegraphics[width=\hsize]{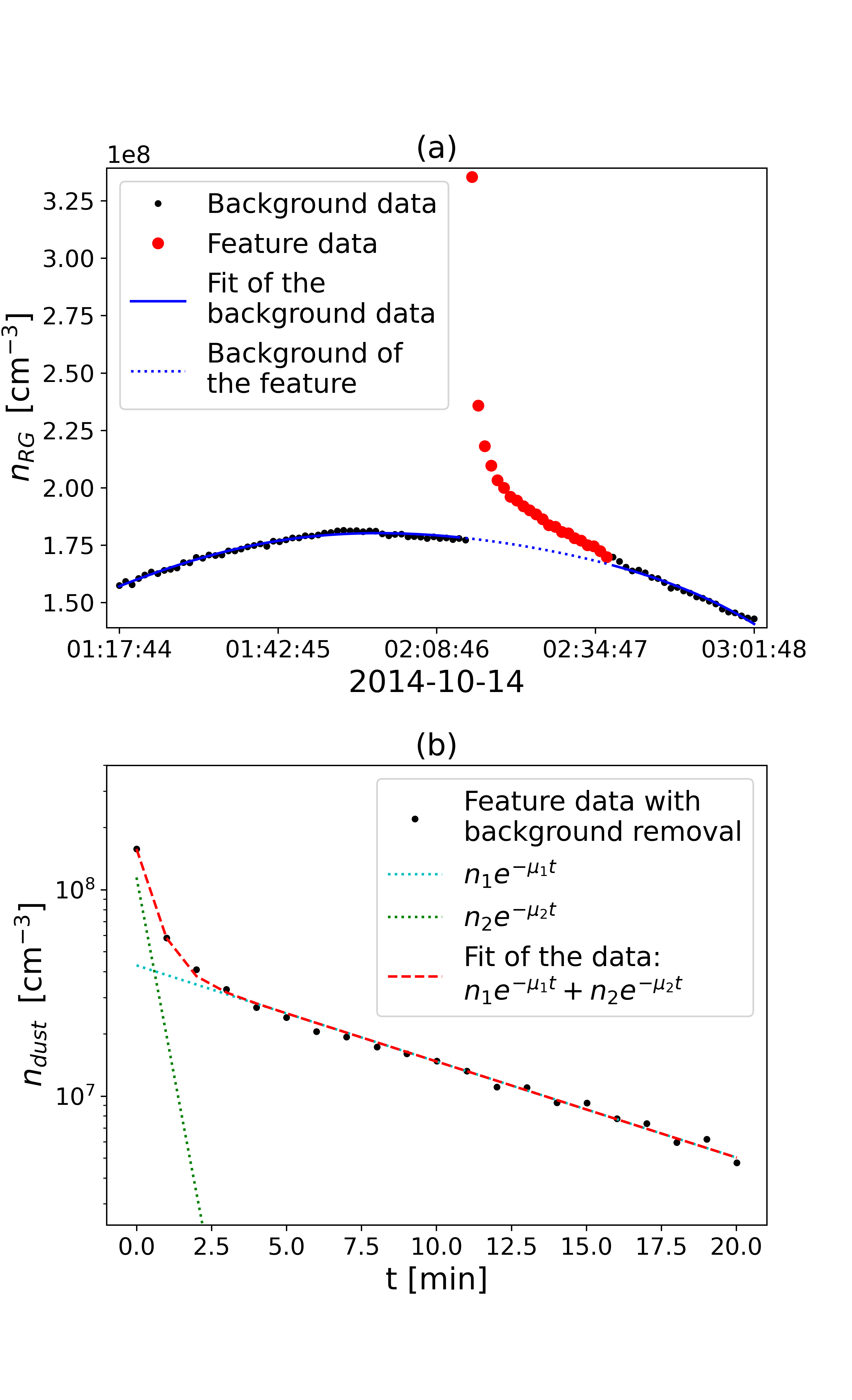}
\caption{Fitting of the RG measurements. The feature chosen is the one on 4 October 2014 at 02:12:46, the same as in Fig.~\ref{fig:motivation}. \textit{Panel a}: density measured by the RG as a function of time. The measurements representing the nominal RG signal (black dots), the data points attributed to the feature (red dots), the fifth degree polynomial fit to the nominal RG signal data (solid blue line), and the extension of the fit to the region where the feature is located (dotted blue line) are displayed. \textit{Panel b}: residuals representing the density due to the outgassing of the volatiles in the icy particle as a function of the time elapsed since the beginning of their sublimation (black dots). The fit to the residuals (red dashed line) is given by the sum of two exponential decay functions (green and light blue dotted lines).}
\label{fig:example_procedure_lambda}
\end{figure}

For every icy particle that has entered the RG and whose sublimating volatiles caused a feature with a time-resolved tail, the identification of the presence of multiple groups making up the volatile part of the detected particle is done by fitting the RG data in the tail. The steps of the interpolation are illustrated in Fig.~\ref{fig:example_procedure_lambda}; the chosen icy particle is the same as Fig.~\ref{fig:motivation} because this particle is the one that shows the least noise of the whole data set. 

First, the boundaries of the feature should be established. This choice can be biased by the incorrect removal or addition of one or more data point at the end of the tail. This may be crucial because an incorrect selection results in a source of error that is difficult to consistently include in the calculation. However, this error is usually small compared to other ones, so it can be neglected. As the analysis only starts from the point with the highest density and subsequently models the decay, also a slow increase -- which is observed for some of the features -- is not critical. One of the reasons for not having a sudden increase in density could be a shower of particles impacting the instrument \citep{Fulle_2015}. The only exception to this assumption occurs when a feature shows that the RG reached saturation, that is, when the gas density presents values higher than the maximal density measurable by the instrument. Saturation leads to a plateau at its top (see for example the first feature measured with science mode in Fig.~\ref{fig:difficulties}a). In this scenario, the tail is extended to compensate for the truncation using the fit derived from RG data points (Eq.~\ref{eqn:fit_decay}).

After the choice of the boundaries, a sufficiently large window is selected around the feature (black dots in Fig.~\ref{fig:example_procedure_lambda}a). Normally, it includes several tens of measurements covering both sides of the feature. The amount of data selected to compose the nominal RG signal has an influence on the results. However, if a sufficiently large range is chosen, this effect could be minimised. The nominal RG signal is fitted with a fifth degree polynomial (solid blue line in Fig.~\ref{fig:example_procedure_lambda}a). The degree of the polynomial was chosen to best represent the nominal RG data points of the coma. A variation analysis of the degree of the polynomial on selected features shows that changing the degree of the polynomial leads to variations in the exponential decay constants of Eq.~\ref{eqn:fit_decay2} below 3\%. However, since the analysis does not require such accuracy, this aspect may be neglected. Each feature we analysed was therefore manually checked for consistency between the fit of the nominal RG signal and the data. The interpolation polynomial is then removed; in this way, the residuals only contain the contribution of the volatile part of the icy particle. As mentioned above, only values found in the tail of the feature are selected (black dots in Fig.~\ref{fig:example_procedure_lambda}b). These values are interpolated, using a least squares method, with the function
\begin{equation}
\label{eqn:fit_decay}
n_\mathrm{dust}(t) = \sum_{i=1}^m n_i\, e^{-\mu_i t},
\end{equation}
which is composed, depending on the feature, of one ($m=1$) or two ($m=2$) exponential decays and, therefore, also one or two exponential decay constants $\mu_i$. In Fig.~\ref{fig:example_procedure_lambda}b, the two exponential decays (green and light blue dotted lines) and the corresponding sum (red dashed line) are shown.

The number of volatile molecules inside the RG at time $t$ is given by
\begin{equation}
\label{eqn:number_particles}
N(t)=n_\mathrm{dust}(t)\,V,
\end{equation}
where $n_{\mathrm{dust}}(t)$
is given by Eq.~\ref{eqn:fit_decay} and $V$
is the inner volume of the RG. The volume is calculated by summing the volume of the spherical cavity and the volume of the cylindrical neck with the subtraction of the volume of the spherical cup of the cavity that lies inside the neck (Fig.~\ref{fig:ram_gauge}). Additionally, the volumes of the microtip field emitter device and of the electron repeller have been neglected. In this step, it is assumed that the density measured in the RG operation region is identical to the one inside the spherical cavity. The differential equation that describes the density inside the RG as a function of time is 
\begin{equation}
\label{eqn:cavity_diff_eqn}
\frac{\mathrm{d}N(t)}{\mathrm{d}t}=Q(t)-F(t),
\end{equation}
where $Q(t)$ is the volatile sublimation rate and $F(t)$ is the flux of outgoing volatiles from the opening of the cavity. The latter quantity can be expressed as
\begin{equation}
\label{eqn:flux}
F(t)=\frac{1}{4}\,n_\mathrm{dust}(t)\,A\,v(t),
\end{equation}
where $A = \pi\cdot(3\,\mathrm{mm})^2$ is the area of the opening of the cavity, \begin{equation}
\label{eqn:velocity}
v(t)=\sqrt{\frac{8\,k_\mathrm{B}\,T_\mathrm{RG}(t)}{\pi\,m}}
\end{equation}
is the velocity of the flux, $k_\mathrm{B}$ is the Boltzmann constant, $T_\mathrm{RG}(t)$ is the temperature of the RG, and $m$ is the mass of a water particle. As water is the most abundant volatile in the cometary coma \citep[e.g.][]{Le_roy_et_al_2015}, our analysis uses water to derive a reference particle volume representing the measured volatile fraction.

The combination and rearrangement of Eqs.~\ref{eqn:number_particles}--\ref{eqn:velocity}, leads to
\begin{equation}
\label{eqn:q_t}
Q(t)=V\,\frac{\mathrm{d}n_\mathrm{dust}(t)}{\mathrm{d}t}+\frac{1}{4}\, n_\mathrm{dust}(t)\, A\sqrt{\frac{8\,k_\mathrm{B}\,T_\mathrm{RG}(t)}{\pi\,m}}.
\end{equation}
Using the fit of Eq.~\ref{eqn:fit_decay} to model the tail, Eq.~\ref{eqn:q_t} can be solved, thereby obtaining the number of volatile molecules that are sublimated over time. This is the starting point that permits both the identification of the number of types of icy particles with different sublimation timescales and the estimation of the volume of volatiles in the icy particles.

Eq.~\ref{eqn:q_t} was fitted, similarly to Eq.~\ref{eqn:fit_decay}, with a least squares method according to
\begin{equation}
\label{eqn:fit_decay2}
Q(t) = \sum_{i=1}^m q_i\,e^{-\lambda_i t},
\end{equation}
either using a single ($m=1$) or a double ($m=2$) exponential decay function. The values of the exponential decay constants of Eq.~\ref{eqn:fit_decay2} represent a density drop proportional to $1/e$ attributable to the loss of material following the sublimation of the volatile components of the icy particles. Since every volatile species -- but also every mixture of volatiles -- has a unique exponential decay constant, by comparing the various $\lambda_i$, it is possible to estimate how many groups of icy particles there are.

The total number of sublimated molecules is given by integrating Eq.~\ref{eqn:fit_decay2} over the whole feature's tail, as
\begin{equation}
\label{eqn:N_sun}
N_\mathrm{sub}=\int_{tail}Q(t)\mathrm{d}t.
\end{equation}
As previously noted, if saturation of the RG occurred, an extension of the tail to compensate for the lack of accurate measurements is performed.
Subsequently, the volume of the sublimated molecules is estimated by multiplying $N_\mathrm{sub}$ by the volume of a $\mathrm{H_2O}$ molecule as $V_\mathrm{sub}=N_\mathrm{sub}\, V_\mathrm{H_2O}$. Since it is not possible to know which molecules make up the icy particles, we decided for simplicity to adopt the liquid water density of 1 g cm$^{-3}$, which can easily be scaled to any other density. Finally, the latter is converted into a diameter, $d$, of an equivalent sphere with the same volume of $V_\mathrm{sub}$, that is
\begin{equation}
\label{eqn:diameter}
d=2\sqrt[3]{\frac{3}{4\,\pi}V_\mathrm{sub}}.
\end{equation}
As for the refractory components of the icy particles and, hence, for the total volume, no conclusion can be drawn because the RG is only capable of detecting volatiles.


\section{Results}
\label{sec:results}
Based on the inspection of the data sets acquired in the two COPS modes (i.e. scientific and monitoring) gas signatures associated with 73 icy particles were identified. It is likely that additional icy particles struck the RG, but for the reasons indicated in Sect.~\ref{subsec:peaks_identification}, these may not be visible in the data.

In Table~\ref{tab:73_aggregates} the features identified are listed according to their start time together with the following information: 1) RG measurement mode; 2) presence or lack of a time-resolved tail; 3) the identification number used to designate features showing a sufficiently time-resolved tail.

\begin{figure}
\centering
\includegraphics[width=\hsize]{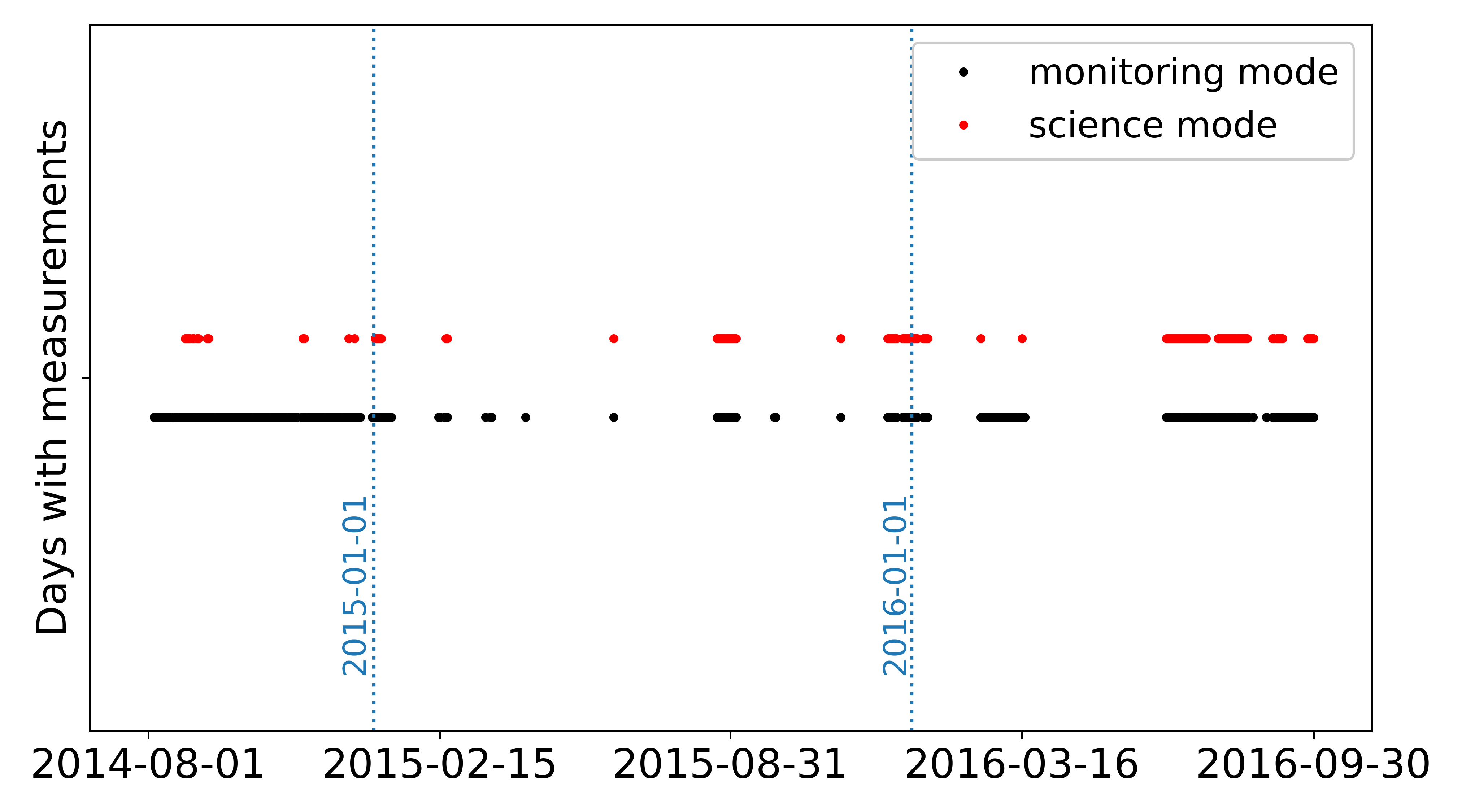}
\caption{Days between 01 August 2014 and 30 September 2016 when the RG carried out measurements (black points for the monitoring mode, red points for the science mode). Large gaps without data, mainly during 2015, are present.}
\label{fig:days_with_meas}
\end{figure}

\begin{table*}
\centering
\caption{List of icy particles found within the RG data.}
\begin{tabular}{lllllllll}
\hline\hline
Start time & Mode & Tail & Id.\ number & $\quad$ & Start time & Mode & Tail & Id.\ number \\
\hline
2014-10-14T02:12:46 & M & Y & 1 & $\quad$ & 2015-08-27T15:25:30 & M & N & \\
2014-10-18T03:32:37 & M & Y & 2 & $\quad$ & 2015-12-18T11:07:58 & M & N & \\
2014-10-19T19:22:18 & M & N & & $\quad$ & 2015-12-27T21:04:00 & M & N & \\
2014-10-20T04:06:38 & M & N & & $\quad$ & 2016-01-11T11:47:51 & M & N & \\
2014-10-21T20:28:22 & M & Y & 3 & $\quad$ & 2016-02-18T14:49:38 & M & N & \\
2014-10-24T23:53:13 & M & N & & $\quad$ & 2016-02-19T12:04:25 & M & N & \\
2014-10-25T18:03:14 & M & N & & $\quad$ & 2016-02-20T21:31:38 & M & N & \\
2014-10-26T09:49:48 & M & N & & $\quad$ & 2016-02-22T14:25:56 & M & Y & 8 \\
2014-10-28T03:30:18 & M & N & & $\quad$ & 2016-02-22T18:03:04 & M & Y & 9 \\
2014-10-28T07:01:27 & M & N & & $\quad$ & 2016-03-01T11:51:33 & M & N & \\
2014-10-28T08:45:31 & M & N & & $\quad$ & 2016-03-08T11:49:19 & M & Y & 10 \\
2014-11-10T15:01:02 & M & Y & 4 & $\quad$ & 2016-03-11T11:26:39 & M & N & \\
2014-11-10T16:38:05 & M & N & & $\quad$ & 2016-03-16T13:13:56 & M & N & \\
2014-11-13T12:40:10 & M & N & & $\quad$ & 2016-03-17T18:18:54 & M & N & \\
2014-11-14T02:09:34 & M & N & & $\quad$ & 2016-07-03T08:13:05 & S & Y & 11 \\
2014-11-14T21:34:25 & M & N & & $\quad$ & 2016-07-03T08:42:06 & S & Y & 12 \\
2014-11-21T05:08:44 & M & N & & $\quad$ & 2016-07-19T02:22:12 & M & N & \\
2014-11-26T04:13:02 & M & Y & 5 & $\quad$ & 2016-07-19T14:13:13 & M & N & \\
2014-11-26T05:52:02 & M & N & & $\quad$ & 2016-07-20T09:30:14 & M & Y & 13 \\
2014-11-26T06:46:02 & M & N & & $\quad$ & 2016-07-22T06:03:51 & M & N & \\
2014-11-26T14:01:02 & M & N & & $\quad$ & 2016-07-30T05:11:06 & M & N & \\
2014-11-28T20:29:28 & M & N & & $\quad$ & 2016-08-06T14:48:19 & M & N & \\
2014-11-29T13:45:15 & M & N & & $\quad$ & 2016-08-12T19:47:10 & M & N & \\
2014-11-29T22:55:33 & M & N & & $\quad$ & 2016-09-05T17:43:43 & S & Y & 14 \\
2014-11-29T23:05:33 & M & N & & $\quad$ & 2016-09-05T17:48:01 & S & Y & 15 \\
2014-11-29T23:41:34 & M & N & & $\quad$ & 2016-09-05T17:48:39 & S & Y & 16 \\
2014-11-30T00:13:35 & M & N & & $\quad$ & 2016-09-05T17:52:27 & S & Y & 17 \\
2014-11-30T02:00:38 & M & N & & $\quad$ & 2016-09-05T17:54:22 & S & Y & 18 \\
2014-12-02T01:29:57 & M & N & & $\quad$ & 2016-09-05T17:55:54 & S & Y & 19 \\
2014-12-04T11:50:10 & M & Y & 6 & $\quad$ & 2016-09-05T17:58:24 & S & Y & 20 \\
2014-12-04T11:59:10 & M & N & & $\quad$ & 2016-09-05T17:59:36 & S & Y & 21 \\
2014-12-04T14:16:13 & M & Y & 7 & $\quad$ & 2016-09-05T18:05:42 & S & Y & 22 \\
2014-12-04T23:19:32 & M & N & & $\quad$ & 2016-09-05T18:09:52 & S & Y & 23 \\
2014-12-07T10:01:49 & M & N & & $\quad$ & 2016-09-05T18:11:14 & S & Y & 24 \\
2014-12-09T14:21:38 & M & N & & $\quad$ & 2016-09-05T18:12:34 & S & Y & 25 \\
2014-12-12T06:44:17 & M & N & & $\quad$ & 2016-09-11T23:38:00 & M & N & \\
2014-12-22T13:33:47 & M & N & & $\quad$ & \\
\hline
\end{tabular}
\tablefoot{In the first column, the start time is shown. In the second column, the mode in which the RG was operating when the detection took place is indicated with S for the scientific mode, and M for the monitoring mode. In the third column, Y indicates that the feature has a time-resolved tail and N that the tail is not visible. Finally, the fourth column shows the identification numbers of the features having a time-resolved tail.}
\label{tab:73_aggregates}
\end{table*}
Of the 73 icy particles, 14 were found in scientific mode. All 14 of their features have a fully time-resolved tail (as compared to only 11 out of the 59 in the monitoring mode). This is due to the fact that with a higher time resolution, even a swift sublimation is measured multiple times. However, the monitoring mode was used more frequently than scientific mode (Fig.~\ref{fig:days_with_meas}), which is why more features were recorded in monitoring mode.

Regarding the time distribution of detections, Table~\ref{tab:features_per_month} shows the number of identified icy particles per month. Notably, between the beginning of August 2014 and the end of September 2016, the RG was operational on only 319 days, mainly at the beginning and at the end of the mission (Fig.~\ref{fig:days_with_meas}).
\begin{table}
\centering
\caption{Number of icy particles detected by the RG per month.}
\begin{tabular}{ll}
\hline\hline
Month & No.\ of features \\
\hline
2014-10 & 11 \\
2014-11 & 17 \\
2014-12 & 9 \\
2015-08 & 1 \\
2015-12 & 2 \\
2016-01 & 1 \\
2016-02 & 5 \\
2016-03 & 5 \\
2016-07 & 7 \\
2016-08 & 2 \\
2016-09 & 13 \\
\hline
\end{tabular}
\label{tab:features_per_month}
\end{table}
However, from the data in Tables~\ref{tab:73_aggregates} and \ref{tab:features_per_month}, we can conclude that most of the detections took place either in the first months of the mission or on 5 September 2016. 
Because the RG emission current and, hence, the RG sensitivity was reduced in December 2015 in order to prevent any excessive degradation of the instrument, no straightforward conclusions can be drawn from the observed detection frequency modulations. Dust activity as a function of the heliocentric distance will be discussed in detail in a follow-up manuscript based on data from the COPS nude gauge (Pestoni et al., in preparation). For the nude gauge, there was no degradation problem and there was full coverage for the mission. The 12 detections on 5 September 2016, on the other hand, occurred during the major outburst described by, for example, \citet{Altwegg_et_al_2017} and \citet{della_corte_2019}. The details of this are given in Sect.~\ref{subsec:features_and_dust}.

In this study, we also investigate the dependence of the results on parameters related to the position of Rosetta during the detection of icy particles. The only significant aspect concerns the cometocentric distance, as highlighted in \citet{della_corte_2019}, wherein the dust flux was analysed. Of the 73 icy particles extracted from the RG measurement data sets, 41\% were measured within 10 km, 92\% within 40 km and 99\% within 100 km from the nucleus surface (during the mission, the cometocentric distance varied between $\sim$3 and 
$\sim$1450 km). There are too few observations available for a robust statistical analysis and the distribution of detected icy particles may be caused by the dilution of the coma due to the expansion of dust and gas. Nevertheless, this could also be taken as an indication that volatile molecules are not present far from the comet because they are sublimated at an earlier point in time. The only exception to this observation is the feature from 27 August 2015, measured more than 400 kilometres away from the comet. This peculiar detection may have been caused by a fragment that retained some of its volatiles longer than the other icy particles. Since this feature does not have a time-resolved tail, it does not affect the results in Sects.~\ref{subsec:amount_volat_speci} and \ref{subsec:size_volat_comp}.

\subsection{Different groups of volatile species}
\label{subsec:amount_volat_speci}
\begin{figure}
\centering
\includegraphics[width=\hsize]{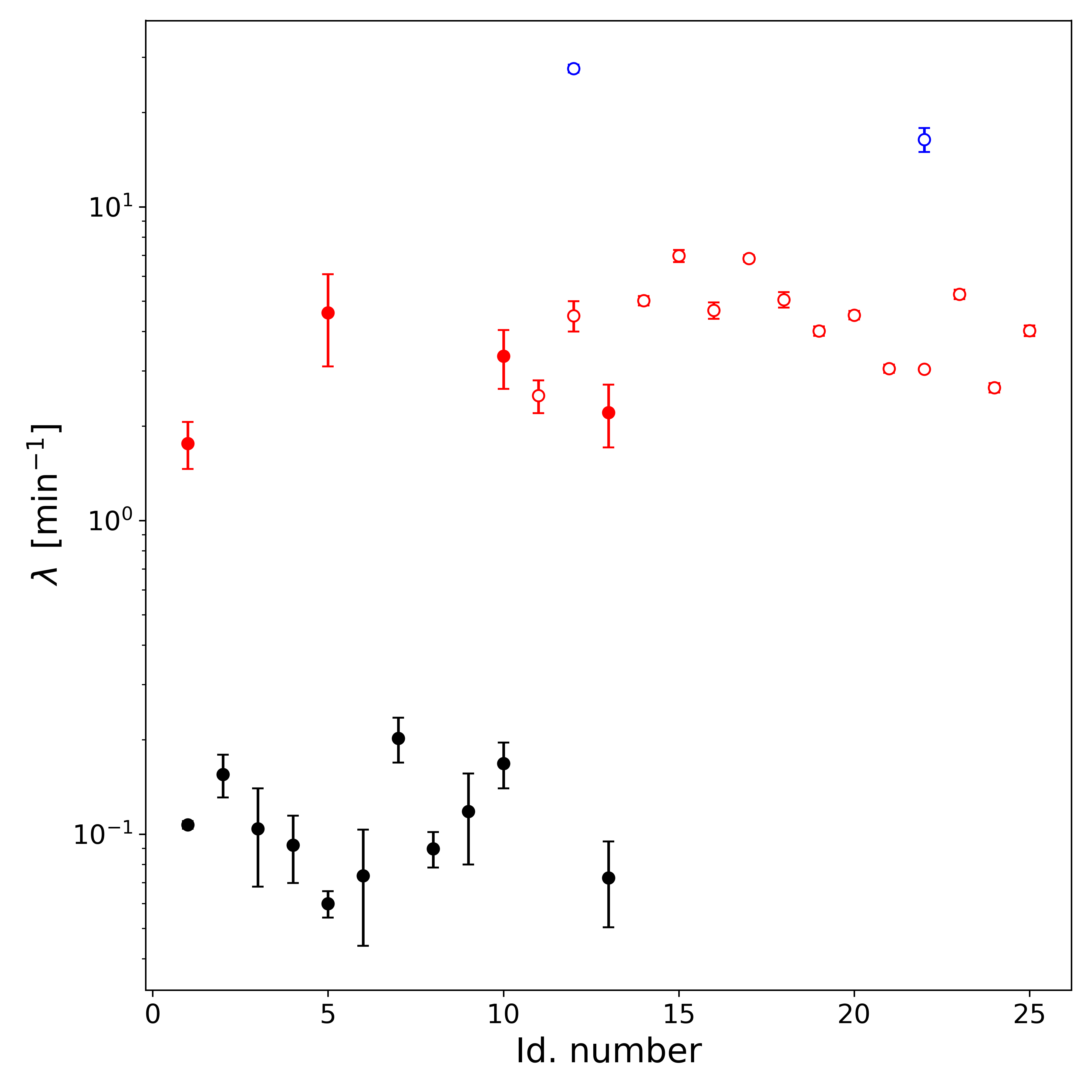}
\caption{Exponential decay constants of the 25 resolved features. Full dots are values corresponding to the monitoring mode, whereas empty dots indicate features extracted from the science mode.}
\label{fig:lambda_plot}
\end{figure}

By applying the procedure depicted in Sect.~\ref{subsec:analysis_procedure} to the 25 features with a time-resolved tail, a plot showing the exponential decay constants of the various icy particles is created (Fig.~\ref{fig:lambda_plot}). The error bars of the data in the plot include the measurement error and the least squares fit error, whereas the error of subtracting the nominal RG signal and the error arising from the number of measurements composing the feature are not considered (Sect.~\ref{subsec:analysis_procedure} for the details).

From Fig.~\ref{fig:lambda_plot}, three distinct populations, with averages $\Bar{\lambda_1}=0.1\,\mathrm{min}^{-1}$ (black points), $\Bar{\lambda_2}=4.1\,\mathrm{min}^{-1}$ (red points), and $\Bar{\lambda_3}=22.0\,\mathrm{min}^{-1}$ (blue points) respectively, are observed. We checked that there is no correlation between these three populations and the quantities associated with Rosetta's position or the temperature of the RG.
The presence of three exponential decay constants means that there are three icy particle groups.
The group indicated by the black dots in the plot, whose volatiles are sublimated more slowly, is present only within the monitoring mode data and includes features where the tail lasts between 5 and 20 minutes. It could not be observed in scientific mode despite its decay time being longer than the two-second time resolution. This is due to data variability and particularly high noise. On the other hand, the most rapidly sublimating type, indicated by the blue dots in the plot, is observed only in icy particles that where detected in scientific mode. The reasoning is the opposite: this component lasts about ten seconds and, therefore, it cannot be identified with the one-minute time resolution of the monitoring mode. The results of this study are consistent with what reported by \citet{Bergantini_Kaiser_2016} and \citet{Altwegg_et_al_2020} regarding the reason why COSIMA did not detect volatile species, but only refractories. At temperatures in the RG between 254 K and 267 K, volatiles are sublimated within a few minutes or even seconds. This is much shorter than the several days at temperatures of 283 K or more that elapsed between the collection of dust particles and their analysis by COSIMA \citep[][]{Fray_et_al_2016}. The computation of theoretical sublimation rates is complex and can be done only if the geometry and the substance used are very well known \citep[see e.g.][]{Tachiwaki_1990}. In the situation considered in this study, it is not possible to perform such a calculation, as there are too many unknown parameters, including the ice thickness, the molecule that compose the volatiles, the initial temperature of the icy particle, and the method of heating the icy particle (e.g. radiation or heat transfer from the RG).

Fig.~\ref{fig:lambda_plot} shows three groups of icy particles. The differences between the groups are related to the differences in the chemical composition of the volatiles (water, carbon dioxide, organics, etc.) in the icy particles. The degree of embedding of volatiles inside the refractory would not lead to discrete values of $\lambda$, but, rather, to a continuum. This means that the observed groups of icy particles have another origin than only the morphology of the cometary particles. 
The chemical composition of the volatiles will be investigated in laboratory experiments and presented in a future work. The objective of such experiments will be to measure the $\lambda$ of common molecules with a ``twin'' instrument identical to the one on board Rosetta.

\subsection{Amount of volatiles in detected icy particles}
\label{subsec:size_volat_comp}
\begin{figure}
\centering
\includegraphics[width=\hsize]{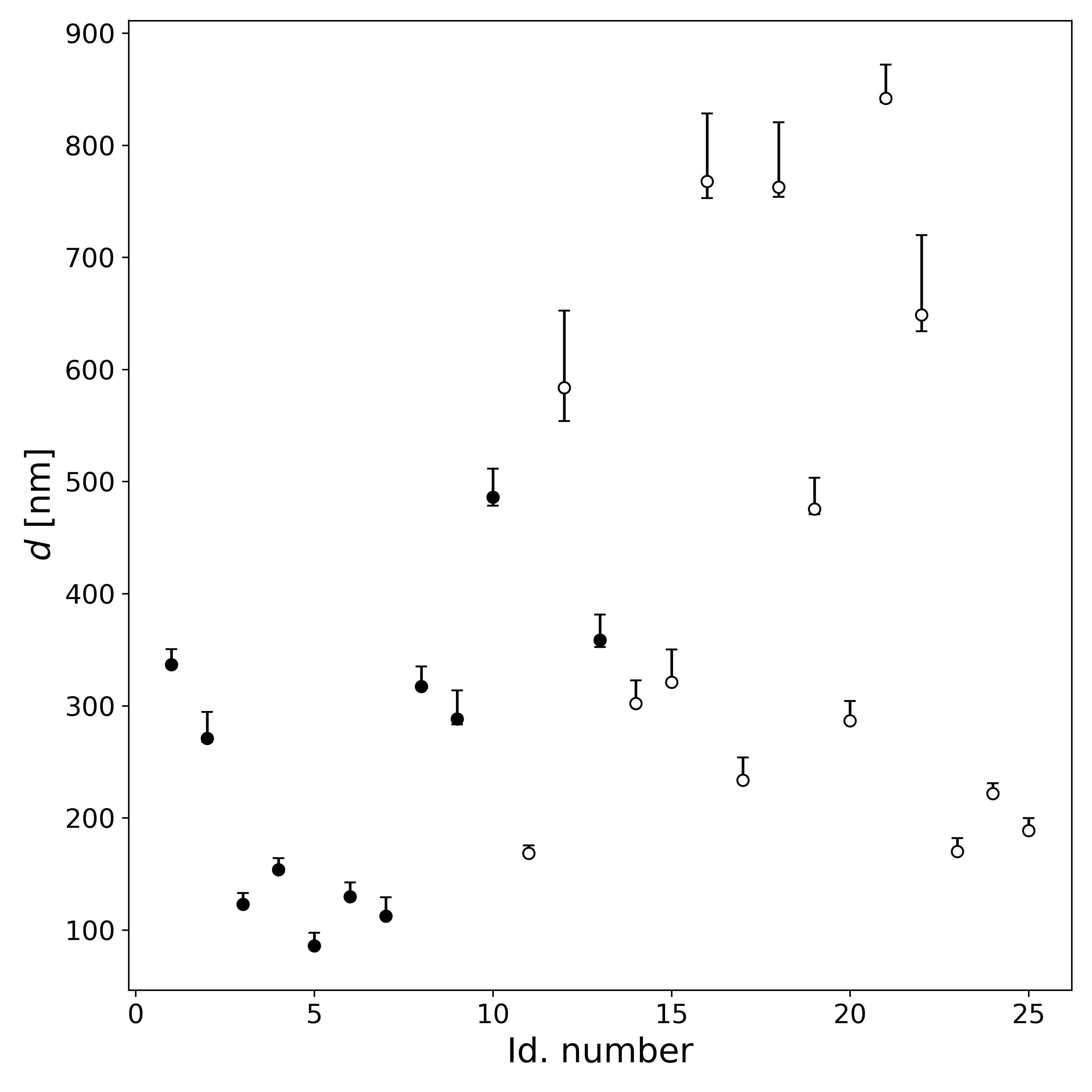}
\caption{Diameters of equivalent water spheres (1 g cm$^{-3}$). As in Fig.~\ref{fig:lambda_plot}, full dots indicate values taken from the monitoring mode, and empty dots are data calculated starting from the science mode features. The diameters obtained here are on the order of hundreds of nanometres.}
\label{fig:diameter_plot}
\end{figure}

\begin{figure}
\centering
\includegraphics[width=\hsize]{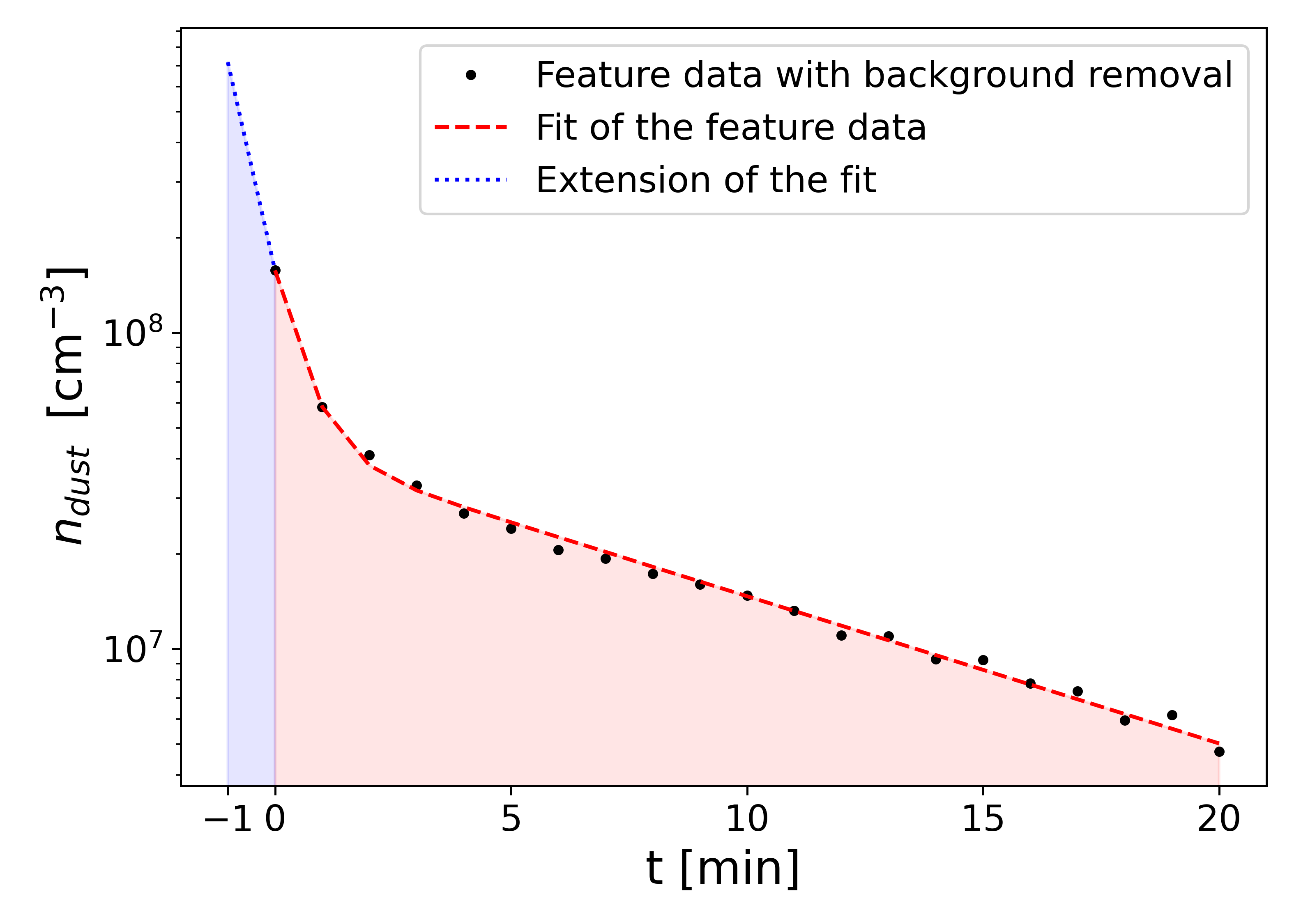}
\caption{Extension of the fit of the feature data. This allows to visualise the source of error on the size of volatiles resulting from the uncertainty in the exact impact time. The chosen feature is the same as the one already proposed in Figs.~\ref{fig:motivation} and \ref{fig:example_procedure_lambda}. 
Since the feature was extracted from the monitoring mode, the fit (red dashed line) on the residuals representing the density due to sublimation (black dots) is extended for sixty seconds (blue dotted line).}
\label{fig:error_procedure}
\end{figure}

Using the fits derived in Sect.~\ref{subsec:amount_volat_speci} and the procedure in Sect.~\ref{subsec:analysis_procedure}, diameters of equivalent spheres containing the same amount of water ice are calculated. The results are shown in Fig.~\ref{fig:diameter_plot}. 
The error bars are based on the error of the fit of the tail, the RG temperature, the fit of the volatile sublimation rate, and the uncertainty with regard to the beginning of the feature due to limited time resolution. It is assumed that the tail starts just after the last measurement before the first data point in the tail. Therefore, the tail's fit must be extended either for two seconds (for the scientific mode), or for sixty seconds (for the monitoring mode), as shown in Fig.~\ref{fig:error_procedure} for the feature already presented in Figs.~\ref{fig:motivation} and \ref{fig:example_procedure_lambda}, to obtain an upper limit for the total volatile content.

The equivalent diameters obtained are on the order of hundreds of nanometres. This range is consistent with the sizes of the grains composing cometary particles found by MIDAS, but for refractories instead of volatiles \citep{Mannel_et_al_2019}. As explained in Sect.~\ref{subsec:peaks_identification}, there may be icy particles whose volatile content is too small to have a sublimation feature that can be extracted from the data because the latter does not stand out sufficiently from the nominal RG signal.


\section{Conclusions}\label{sec:conclusion}
This paper presents a first study on the identification and categorisation of 73 features in RG measurements. Based on the outburst on 5 September 2016, it was possible to infer that these features are the result of the sublimation of volatiles contained in icy particles originating from 67P. Some of the features could have been missed by this analysis due to the noise in the RG signal, and the measurements do not cover the full mission duration.

We identified 12 features related to the outburst on 5 September 2016 with which we are able to show that at least some of the icy particles reaching Rosetta during this event had a volatile component. The COPS RG observations show that the equivalent size of the detected particles' volatile content is on the order of hundreds of nanometres. The results of this study are, therefore, complementary to those obtained by GIADA, MIDAS, and COSIMA since these instruments investigated the refractory component of dust, whereas the RG observed volatiles.

From fitting 25 time-resolved features and comparing their sublimation characteristics, we found three distinct groups of icy particles. The explanation for the observation of different groups is a different composition of the sublimating component of the icy particles. A next step could be an attempt to relate the volatiles in the icy particles observed by the RG to the dust composition obtained by DFMS \citep[][]{Altwegg_et_al_2017,Schuhmann_et_al_2019,Altwegg_et_al_2020}.

Despite some caveats, such as limited coverage throughout the mission, noise from the nominal RG signal, and unknown contributions from refractories, this work demonstrates that the RG can be used to detect the sublimation of volatiles in icy particles. One approach for distinguishing icy particles by the sublimation timescale of their volatile content is also presented here.
Other critical aspects of the COPS measurements include the composition of the volatiles and morphology and total size of the icy particles.

Future research will focus on the analysis of data from the nude gauge to search for dust impacts and associated laboratory measurements. Connections between the two gauges of COPS will be investigated and the results will be compared to other instruments of the Rosetta mission.


\begin{acknowledgements}
Work at the University of Bern was funded by the State of Bern, the Swiss National Science Foundation (200020\textunderscore182418), and the European Space Agency through the Rosetta data fusion: Dust and gas coma modelling (5001018690) grant. The results from ROSINA would not be possible without the work of the many engineers, technicians, and scientists involved in the mission, in the Rosetta spacecraft, and in the ROSINA instrument team over the past 20 years, whose contributions are gratefully acknowledged. Rosetta is an European Space Agency (ESA) mission with contributions from its member states and NASA. We thank herewith the work of the whole ESA Rosetta team. All ROSINA flight data have been be released to the PSA archive of ESA and to the PDS archive of NASA.

We also thank the International Space Science Institute (ISSI) team ``Characterization Of Cometary Activity Of 67P/Churyumov-Gerasimenko Comet''
for valuable discussion and helpful suggestions.

We warmly thank the referee and the editor for constructive comments that contributed to vastly improve our article. 
\end{acknowledgements}

\bibliographystyle{aa} 
\bibliography{ram_gauge} 

\end{document}